\documentclass[12pt]{article}
\usepackage{graphicx}

\def\downstrut{\vrule height 1ex depth 3.0ex width 0pt}
\def\medstrut{\vrule height 1ex depth 1.0ex width 0pt}

\def \beq{\begin{equation}}
\def \eeq{\end{equation}}
\def\eqref#1{(\ref{#1})}
\def\bea{\begin{eqnarray}}
\def\eea{\end{eqnarray}}

\def\red{
}
\def\black{
}

\def\URLtilde{\lower0.2em\hbox{$\tilde{\phantom{a}}$}}
\def\mycomm#1{\hfill\break\strut\kern-3em{\red\tt ====> #1\black}\hfill\break}
\def\mycommNL#1{\strut\kern-3em{\red\tt ====> #1\black}\hfill\break}

\begin{document}
\thispagestyle{empty}
\rightline{EFI 08-09}
\rightline{TAUP 2875/08}
\rightline{WIS/07/08-Apr-DPP}
\rightline{ANL-HEP-PR-08-20}
\rightline{arXiv:0804.1575}
\vskip1.5cm

\centerline{\large \bf The Quark Model and $b$ Baryons}
\bigskip

\centerline{Marek Karliner$^a$, Boaz Keren-Zur$^a$, Harry J. Lipkin$^{a,b,c}$,
and Jonathan L. Rosner$^d$}
\medskip

\centerline{$^a$ {\it School of Physics and Astronomy}}
\centerline{\it Raymond and Beverly Sackler Faculty of Exact Sciences}
\centerline{\it Tel Aviv University, Tel Aviv 69978, Israel}
\medskip

\centerline{$^b$ {\it Department of Particle Physics}}
\centerline{\it Weizmann Institute of Science, Rehovoth 76100, Israel}
\medskip

\centerline{$^c$ {\it High Energy Physics Division, Argonne National
Laboratory}}
\centerline{\it Argonne, IL 60439-4815, USA}
\medskip

\centerline{$^d$ {\it Enrico Fermi Institute and Department of Physics}}
\centerline{\it University of Chicago, 5640 S. Ellis Avenue, Chicago, IL
60637, USA}
\bigskip
\strut

\centerline{\bf ABSTRACT}
\bigskip

\begin{quote}
The recent observation at the Tevatron of $\Sigma_b^{\pm}$ ($uub$ and $ddb$) 
baryons within 2 MeV of the predicted $\Sigma_b - \Lambda_b$ splitting
and of $\Xi_b^-$ $(dsb)$ baryons at the Tevatron within a few MeV of
predictions has provided strong confirmation for a theoretical
approach based on modeling the color hyperfine interaction.  The prediction of
$M(\Xi^-_b) = 5790$ to 5800 MeV is reviewed and similar methods used to predict
the masses of the excited states $\Xi_b^\prime$ and $\Xi_b^*$.  The main
source of uncertainty is the method used to estimate the mass difference
$m_b - m_c$ from known hadrons.  We verify that corrections due to the details
of the interquark potential and to $\Xi_b$--$\Xi_b^\prime$ mixing are small.
For S-wave $qqb$ states we predict $M(\Omega_b) = 6052.1 \pm 5.6$ MeV,
$M(\Omega^*_b) = 6082.8 \pm 5.6$ MeV, and $M(\Xi_b^0) = 5786.7 \pm 3.0$ MeV.
For states with one unit of orbital angular momentum between the $b$ quark
and the two light quarks we predict $M(\Lambda_{b[1/2]}) = 5929 \pm 2$ MeV,
$M(\Lambda_{b[3/2]}) = 5940 \pm 2$ MeV, $M(\Xi_{b[1/2]}) = 6106 \pm 4$ MeV,
and $M(\Xi_{b[3/2]}) = 6115 \pm 4$ MeV.  Results are compared with those of
other recent approaches.
\end{quote}
\bigskip

\leftline{PACS codes: 14.20.Mr, 12.40.Yx, 12.39.Jh, 11.30.Hw}

\newpage

\begin{section}{Introduction}

The first observed baryon with a $b$ quark was the isospin-zero $\Lambda_b$,
whose mass has recently been well-measured: $M(\Lambda_b) = 5619.7 \pm
1.2 \pm 1.2$ MeV \cite{Acosta:2005mq}.  Its quark content is $\Lambda_b = bud$,
where the $ud$ pair has spin and isospin $S(ud) = I(ud) =0$.  Now the CDF 
Collaboration has observed candidates for $\Sigma_b^\pm$ and $\Sigma_b^{*\pm}$
\cite{Aaltonen:2007rw} with masses consistent with predictions \cite{earlier,%
Jenkins:1996de,Karliner:2003sy,Ebert:2005xj,Rosner:2006yk,Karliner:2006ny,%
Liu:2007fg,Roberts:2007ni,Jenkins:2007dm}.  D0 and CDF have seen candidates for
$\Xi_b^- = bsd$ \cite{Abazov:2007ub,Aaltonen:2007un}.  The more precise CDF
mass lies close to a prediction based on careful accounting for wave function
effects in the hyperfine interaction \cite{v1}.

The CDF sensitivity appears adequate to detect further heavy baryons.  The
S-wave levels of states containing $bsu$ or $bsd$ consist of the
$J=1/2$ states $\Xi_b^{0,-}$ and ${\Xi'}_b^{(0,-)}$ and the $J=3/2$ states
$\Xi_b^{*(0,-)}$.  Additional baryonic states containing the $b$ quark
include $\Omega_b = bss~(J=1/2)$, $\Omega^*_b = bss~(J=3/2)$, and orbital
excitations of $\Lambda_b$ and other $b$-flavored baryons.  In this paper we
predict the masses of these states and estimate the dependence of the
predictions on the form of the interquark potential, extending a previous
application to hyperfine splittings of known heavy hadrons
\cite{Keren-Zur:2007vp}.  Two observations based on a study of the hadronic
spectrum lead to improved predictions for the $b$ baryons.  The first is that
the effective mass of the constituent quark depends on the spectator quarks
\cite{Karliner:2003sy}, and the second is an effective supersymmetry
\cite{Karliner:2006ny} -- a resemblance between mesons and baryons where the
anti-quark is replaced by a diquark \cite{Lichtenberg:1989ix}.
Parts of this article have appeared previously in preliminary form \cite{v1,%
Karliner:2007cu}.

We review the predictions for $\Sigma_b$ and $\Sigma^*_b$ in Section 2, and
discuss predictions for $M(\Xi_b)$ in Section 3, starting with an
extrapolation from $M(\Xi_c)$ without correction for hyperfine (HF)
interaction and then estimating this correction.  In the $\Xi_b$ the light
quarks are approximately in a state with $S=0$, while another heavier state
${\Xi'}_b$ is expected in which the light quarks mainly have $S=1$. There is
also a state $\Xi_b^*$ expected with light-quark spin 1 and total $J=3/2$.
Predictions for ${\Xi'}_b$ and $\Xi_b^*$ masses are discussed in Section 4.
We estimate the effect of mixing between light-quark spins $S=0$ and 1 in
Section 5, and isospin splittings of the $\Xi_b$ family of states in Section 6.
Section 7 is devoted to predictions of $M(\Omega_b)$ and $M(\Omega^*_b)$, while
Section 8 treats orbital excitations.  Comparisons with other approaches are
made in Section 9, while Section 10 summarizes.
\end{section}

\begin{section}{The $\Sigma_b$ and $\Sigma_b^{*\pm}$ states}

The $\Sigma_b^\pm$ states consist of a light quark pair $uu$ or $dd$ (a
``nonstrange diquark'') with $S=I=1$ coupled with the $b$ quark to $J=1/2$,
while in the $\Sigma_b^{*\pm}$ states the light quark pair and the $b$ quark
are coupled to $J=3/2$.  The corresponding $ud$ pair in the $\Lambda_b$ has
$S=I=0$.  The experimental $\Sigma_b$--$\Lambda_b$ mass differences
\cite{Aaltonen:2007rw},

\bea
M(\Sigma_b^-) - M(\Lambda_b) = 195.5^{+1.0}_{{-}1.0}\,({\rm stat.}) \pm
0.1\, \hbox{(syst.) MeV}
\nonumber\\
\\
M(\Sigma_b^+) - M(\Lambda_b) = 188.0^{+2.0}_{{-}2.3}\,({\rm stat.}) \pm
0.1\, \hbox{(syst.) MeV}
\nonumber
\eea
with isospin-averaged mass difference $M(\Sigma_b) - M(\Lambda_b) = 192$ MeV,
are to be compared with the prediction \cite{Karliner:2003sy,Karliner:2006ny}
$M_{\Sigma_b} - M_{\Lambda_b} = 194 \,{\rm MeV}$.  Note also:

(1) The mass difference between spin-1 and spin-zero nonstrange diquarks
governs the splitting between the spin-weighted average $[2 M(\Sigma_b^*) +
M(\Sigma_b)]/3$ and the $\Lambda_b$,
\beq
\frac{M(\Sigma_b) + 2 M(\Sigma_b^*)}{3} - M(\Lambda_b) = (205.9 \pm 1.8)~{\rm
MeV},~
\eeq
where we have used the averages of the differences for $\Sigma_b^{(*)\pm}$.
This should be the same as the corresponding quantity for charmed baryons,
\beq
\frac{M(\Sigma_c) + 2 M(\Sigma_c^*)}{3} - M(\Lambda_c) = (210.0 \pm 0.5)~{\rm
 MeV},
\eeq
and that for strange baryons,
\beq
\frac{M(\Sigma) + 2 M(\Sigma^*)}{3} - M(\Lambda) = (205.1 \pm 0.3)~{\rm MeV}~,
\eeq
where the masses are from Ref.\ \cite{Yao:2006px}, and an
average over the $\Sigma$ isospin multiplet is taken.  In each case the
dominant source of error is the mass of the $I_3 = 0$, $J=3/2$ state,
$\Sigma_c^{*+}$ or $\Sigma^{*0}$.  The agreement is quite satisfactory.

(2) The charge-averaged hyperfine splitting between the $J=1/2$ and $J=3/2$
states involving the spin-1 nonstrange diquark may be predicted from that for
charmed particles:
\beq \label{eqn:hfratio}
\frac{M(\Sigma_b^*) - M(\Sigma_b)}{M(\Sigma_c^*) - M(\Sigma_c)} =
\frac{m_c}{m_b} = \frac{1.5~{\rm GeV}}{4.9~{\rm GeV}} = 0.31~~,
\eeq
where ``constituent'' quark masses are from Ref.\ \cite{Kwong:1987ak}.
Using isospin-averaged differences $M(\Sigma_c) - M(\Lambda_c) = (167.09
\pm 0.13)$ MeV and $M(\Sigma^*_c) - M(\Lambda_c) = (231.5 \pm 0.8)$ MeV
\cite{Yao:2006px}, we predict $M(\Sigma_b^*) - M(\Sigma_b) = 20.0 \pm 0.3$
MeV.  This agrees with the observed splitting (see Table \ref{tab:sigb}).

In analyzing their data on $\Sigma_b^\pm$ and $\Sigma_b^{*\pm}$, the CDF
Collaboration assumed equal mass splittings $M(\Sigma_b^-) - M(\Sigma_b^+)$
and $M(\Sigma_b^{*-}) - M(\Sigma_b^{*+})$.  This assumption was found to be
valid within the experimental errors.  In Ref.\ \cite{Rosner:2006yk} 
a relation $\Sigma^*_{b1} - \Sigma_{b1} = 0.40 \pm 0.07$ MeV was proved between
the $\Delta I = 1$ mass differences $\Sigma_{b1} \equiv M(\Sigma_b^+) -
M(\Sigma_b^-)$ and $\Sigma^*_{b1} \equiv M(\Sigma_b^{*+}) - (\Sigma_b^{*-})$.

\begin{table}[h]
\caption{Values of $Q^{(*)\pm} \equiv M(\Sigma_b^{(*)\pm}) - M(\pi^\pm) -
M(\Lambda_b)$ and $M(\Sigma^{(*)\pm})$ \cite{Aaltonen:2007rw}.
\label{tab:sigb}}
\begin{center}
\begin{tabular}{c c c} \hline \hline
             & $Q^\pm$ or $Q^* - Q$ &     Mass    \\
State        &        (MeV)         &     (MeV)   \\ \hline
$\Sigma_b^+$ & $48.5^{+2.0+0.2}_{-2.2-0.3}$ & $5807.8^{+2.0}_{-2.2}\pm 1.7$ \\
$\Sigma_b^-$ &    $55.9 \pm 1.0 \pm 0.2$    & $5815.2 \pm 1.0 \pm 1.7$    \\
$\Sigma_b^{*+}$ & $21.2^{+2.0+0.4}_{-1.9-0.3}$
 & $5829.0^{+1.6+1.7}_{-1.8-1.8}$ \\
$\Sigma_b^{*-}$ & & $5836.4 \pm 2.0^{+1.8}_{-1.9}$ \\ \hline \hline
\end{tabular}
\end{center}
\end{table}

\end{section}

\begin{section}{$\Xi_b$ mass prediction}
In our model the mass of a hadron is given by the sum 
of the constituent quark masses plus the color-hyperfine (HF) interactions:
\begin{equation}
V^{HF}_{ij}=v\frac{\vec{\sigma_i}\cdot\vec{\sigma_j}}{m_im_j}\langle\delta(r_{ij})\rangle
\end{equation}
where the $m_i$ is the mass of the $i$'th constituent quark, $\sigma_i$ its
spin, $r_{ij}$ the distance between the quarks and $v$ is the interaction
strength. We shall neglect the mass differences between $u$ and $d$
constituent quarks, writing $u$ to stand for either $u$ or $d$. All the hadron
masses (the ones used and the predictions) are for isospin-averaged baryons
and are given in MeV.

The $s$ and $u$ quarks in $\Xi_q$ ($q$ standing for $c$ or $b$) are assumed to
be in relative spin 0 and the total mass is given by the expression:
\begin{eqnarray}
\Xi_q=m_q+m_s+m_u-\frac{3v\langle \delta(r_{us}) \rangle}{m_um_s}
\end{eqnarray}
The $\Xi_b$ mass can thus be predicted using the known $\Xi_c$ baryon mass as a
starting point and adding the corrections due to mass differences and HF
interactions:
\begin{eqnarray}
\Xi_b&=&\Xi_c + (m_b - m_c) -\frac{3v}{m_um_s}\Bigg( \langle \delta(r_{us})
\rangle_{\Xi_b} -   \langle \delta(r_{us}) \rangle_{\Xi_c} \Bigg)
\end{eqnarray}

The observed masses for the charmed-strange baryons $\Xi_c$,
$\Xi'_c$, and $\Xi^*_c$ are \cite{Yao:2006px}:
\begin{equation}
\Xi_c=2469.5\pm0.5~\rm{MeV} \qquad
\Xi_c'=2577\pm4~\rm{MeV} \qquad
\Xi_c^*=2646.3\pm1.8~\rm{MeV} ~.
\end{equation}
\begin{subsection}{Constituent quark mass difference}
The mass difference $(m_b - m_c)$ can be obtained from experimental data using
one of the following expressions:

\begin{itemize}

\item
We can simply take the difference of the masses of the $\Lambda_q$ baryons,
ignoring the differences in the HF interaction:
\begin{eqnarray}
m_b - m_c = \Lambda_b - \Lambda_c = (3333.2 \pm 1.2)~{\rm MeV}~.
\label{eq_lambda_b_lambda_c}
\end{eqnarray}

\item 
We can use the spin averaged masses of the $\Lambda_q$ and $\Sigma_q$ baryons:
\begin{eqnarray}
m_b - m_c = \Bigg(\frac{2 \Sigma_b^* + \Sigma_b+ \Lambda_b}{4}
- \frac{2\Sigma_c^* + \Sigma_c + \Lambda_c}{4}\Bigg) = (3330.4 \pm 1.8)
~{\rm MeV}~.
\label{eq_sigma_b_sigma_c}
\end{eqnarray}

\item 
Since the $\Xi_q$ baryon has strangeness 1, it might be better to use masses of
mesons with $S=1$:
\begin{eqnarray}
m_b - m_c = \Bigg(\frac{3B_s^* + B_s}{4} - \frac{3D_s^* + D_s}{4}\Bigg)
 = (3324.6 \pm 1.4)~{\rm MeV}~.
\label{eq_B_s_D_s}
\end{eqnarray}

\end{itemize}
\end{subsection}

\begin{subsection}{HF interaction correction}

The HF interaction correction can also be based on $\Xi_c$ baryon experimental
data:
\begin{eqnarray}
\frac{v}{m_um_s}\Bigg( \langle \delta(r_{us}) \rangle_{\Xi_b} - \langle
\delta(r_{us}) \rangle_{\Xi_c} \Bigg)
&=&\frac{v\langle \delta(r_{us}) \rangle_{\Xi_c}}{m_um_s}\Bigg(\frac{\langle
\delta(r_{us}) \rangle_{\Xi_b}}{\langle \delta(r_{us}) \rangle_{\Xi_c}}-1
\Bigg) \nonumber \\
=\frac{2\Xi_c^*+\Xi_c'-3\Xi_c}{12}\Bigg(\frac{\langle \delta(r_{us})
\rangle_{\Xi_b}}{\langle \delta(r_{us}) \rangle_{\Xi_c}}-1 \Bigg)
&=&\Bigg(\frac{\langle \delta(r_{us}) \rangle_{\Xi_b}}{\langle \delta(r_{us})
\rangle_{\Xi_c}}-1 \Bigg)(38.4\pm0.5)~\rm{MeV}
\label{eq_HF_correction}
\end{eqnarray}
This expression requires the calculation of the $\delta$ function expectation
value using 3-body wavefunctions from a variational method
\cite{Keren-Zur:2007vp}.  One only needs the shape of the confining potential,
as coupling constants cancel out in the ratio of the $\delta$
function expectation values.  The potentials considered here are the linear,
Coulomb and Cornell (Coulomb + linear) potentials.  We also note results
obtained without the HF corrections.  For the Cornell potential we have an
additional parameter determining the ratio between the strengths of the linear
and Coulombic parts of the potential.  In these calculations we used the
parameters extracted in \cite{Cornell:1980} from analysis of quarkonium spectra
(or $K=0.45$ in the parametrization of \cite{Keren-Zur:2007vp}).

As a test case we compared the values obtained from experimental data and
variational calculations for the ratio of contact probabilities in $\Xi$ and
$\Xi_c$:
\begin{eqnarray}
\frac{2\Xi_c^*+\Xi_c'-3\Xi_c}{2(\Xi^*-\Xi)}=
\frac{\displaystyle{\frac{6v\langle \delta(r_{us})\rangle_{\Xi_c}}{m_um_s}}}
{\displaystyle{\frac{6v\langle \delta(r_{us})\rangle_{\Xi}}{m_um_s}}}=
\frac{\langle \delta(r_{us})\rangle_{\Xi_c}}{\langle \delta(r_{us})\rangle_{\Xi}}
\label{eq_Xi_Xic}
\end{eqnarray}
The results in Table \ref{tab_Xi_Xic} show good agreement between data
and theoretical predictions using the Cornell potential.

\begin{table}
\caption{Comparison between experimental data and predictions of the
ratio of $u$ and $s$ contact probabilities in $\Xi$ and $\Xi_c$
(Eq.\ (\ref{eq_Xi_Xic})).
\label{tab_Xi_Xic}}
\begin{center}
\begin{tabular}{ccc} \hline \hline
 &${\langle \delta(r_{us})\rangle_{\Xi_c}}/{\langle
 \delta(r_{us})\rangle_{\Xi}}$ \\ \hline 
Experimental data~\cite{Yao:2006px} & $1.071\pm0.069$ \\ \hline
Linear                 & $1.022\pm0.072$ \\
Coulomb                & $1.487\pm0.002$ \\ 
Cornell                & $1.063\pm0.047$ \\ \hline \hline
\end{tabular}
\end{center}
\end{table}

\end{subsection}

\begin{subsection}{Results}

The predictions for $M(\Xi_b)$ under various assumptions about constituent
quark mass differences and confinement potentials are given in Table
\ref{tab_Xib}.  In Ref.\ \cite{Keren-Zur:2007vp} we find that the Coulomb
potential leads to a very strong dependence on quark masses not seen in the
data, so one should give these predictions less weight. Ignoring the Coulomb
potential, one gets a prediction for $M(\Xi_b)$ in the range 5790--5800 MeV.

The predictions of Table \ref{tab_Xib} were first presented in Ref.\ \cite{v1}.
At that time we learned of the $\Xi_b^-$ observation
in the $J/\psi \Xi^-$ decay mode by the D0
Collaboration \cite{Abazov:2007ub}.  Subsequently the CDF Collaboration
released their very precise measurement of $M(\Xi_b^-)$ in the same decay
channel \cite{Aaltonen:2007un}.  The reported masses, Gaussian widths (due to
instrumental resolution), and significances of the signal are summarized in
Table~\ref{tab:xib_exp} and in Fig.\ \ref{fig:comp}.  CDF
also sees a significant $\Xi_b^- \to \Xi_c^0 \pi^-$ signal with mass consistent
with that found in the $J/\psi \Xi^-$ mode.

\begin{table}
\caption{Predictions for the $\Xi_b$ mass with various confining
potentials and methods of obtaining the quark mass difference $m_b-m_c$.
\label{tab_Xib}}
\begin{center}
\begin{tabular}{cccc} \hline \hline
$m_b-m_c =$ & $\Lambda_b-\Lambda_c$ 
            & ${\Sigma_b}-{\Sigma_c}$ 
            & ${B_s}-{D_s}$          \\  
& Eq.~(\ref{eq_lambda_b_lambda_c})
& Eq.~(\ref{eq_sigma_b_sigma_c})& Eq.~(\ref{eq_B_s_D_s}) \\ \hline
No HF   correction & $5803\pm2$   & $5800\pm2$  & $5794\pm2$ \\
Linear             & $5801\pm11$  & $5798\pm11$ & $5792\pm11$ \\
Coulomb            & $5778\pm2$   & $5776\pm2$  & $5770\pm2$ \\
Cornell            & $5799\pm7$   & $5796\pm7$  & $5790\pm7$ \\
\hline \hline
\end{tabular}
\end{center}
\end{table}

\begin{table}
\caption{Observations of $\Xi_b^- \to J/\psi \Xi^-$ at the
Fermilab Tevatron.  Errors on mass are (statistical, systematic).
\label{tab:xib_exp}}
\begin{center}
\begin{tabular}{c c c} \hline \hline
  & D0 \cite{Abazov:2007ub} & CDF \cite{Aaltonen:2007un} \\ \hline
Mass (MeV) & $5774 \pm 11 \pm 15$ & $5792.9 \pm 2.5 \pm 1.7$ \\
Width (MeV) & $37 \pm 8$ & $\sim 14$ \\
Significance & $5.5 \sigma$ & $7.8 \sigma$ \\ \hline \hline
\end{tabular}
\end{center}
\end{table}

The D0 mass is consistent with all our predictions for the isospin-averaged
mass, while that of CDF allows us to rule out the (previously disfavored
\cite{Keren-Zur:2007vp}) prediction based on the Coulomb potential.  Both
experiments also agree with a prediction in Ref.\ \cite{Jenkins:1996de},
$M(\Xi_b) = M(\Lambda_b) + (182.7 \pm 5.0)$ MeV $ = (5802.4 \pm 5.3)$ MeV,
where differences in wave function effects were not discussed and $m_b-m_c$ was
taken from baryons only.  (Here we have updated the prediction of
Ref.~\cite{Jenkins:1996de} using the recent CDF \cite{Acosta:2005mq} value of
$M(\Lambda_b)$.) In our work the
optimal value of  $m_b-m_c$ was obtained from $B_s$ and $D_s$ mesons which
contain both heavy and strange quarks, as do $\Xi_b$ and $\Xi_c$.  See also
Refs.~\cite{Ebert:2005xj} and \cite{Liu:2007fg} for compilations of earlier
predictions for the $\Xi_b$ mass; we shall return to this question in Sec.\ 9.
The dependence of $m_b-m_c$ obtained from $B$ and $D$ mesons upon the flavor of
the spectator quark was noted in Ref.~\cite{Karliner:2003sy} where Table I
shows that the value is the same for mesons and baryons not containing strange
quarks but different when obtained from $B_s$ and $D_s$ mesons.

\end{subsection}
\end{section}

\begin{section}{$\Xi_b^*$, $\Xi_b'$ mass prediction}
\begin{subsection}{Spin averaged mass (2$\Xi_b^*+\Xi_b'$)/3}
The $s$ and $u$ quarks of the $\Xi_q^*$ and $\Xi_q'$ baryons are assumed to be
in a state of relative spin 1.  We then find
\begin{eqnarray}
\Xi_q^*&=&m_q+m_s+m_u+v\Bigg( \frac{\langle \delta(r_{qs}) \rangle}{m_qm_s}
       +\frac{\langle \delta(r_{qu}) \rangle}{m_qm_u}
       +\frac{\langle \delta(r_{us}) \rangle}{m_um_s}\Bigg)\\ \nonumber
\Xi_q'&=&m_q+m_s+m_u+v\Bigg( \frac{-2\langle \delta(r_{qs}) \rangle}{m_qm_s}
       +\frac{-2\langle \delta(r_{qu}) \rangle}{m_qm_u}
       +\frac{\langle \delta(r_{us}) \rangle}{m_um_s}\Bigg)\\ \nonumber
\end{eqnarray}
\begin{figure}
\begin{center}
\includegraphics[width=9.0cm,angle=90]{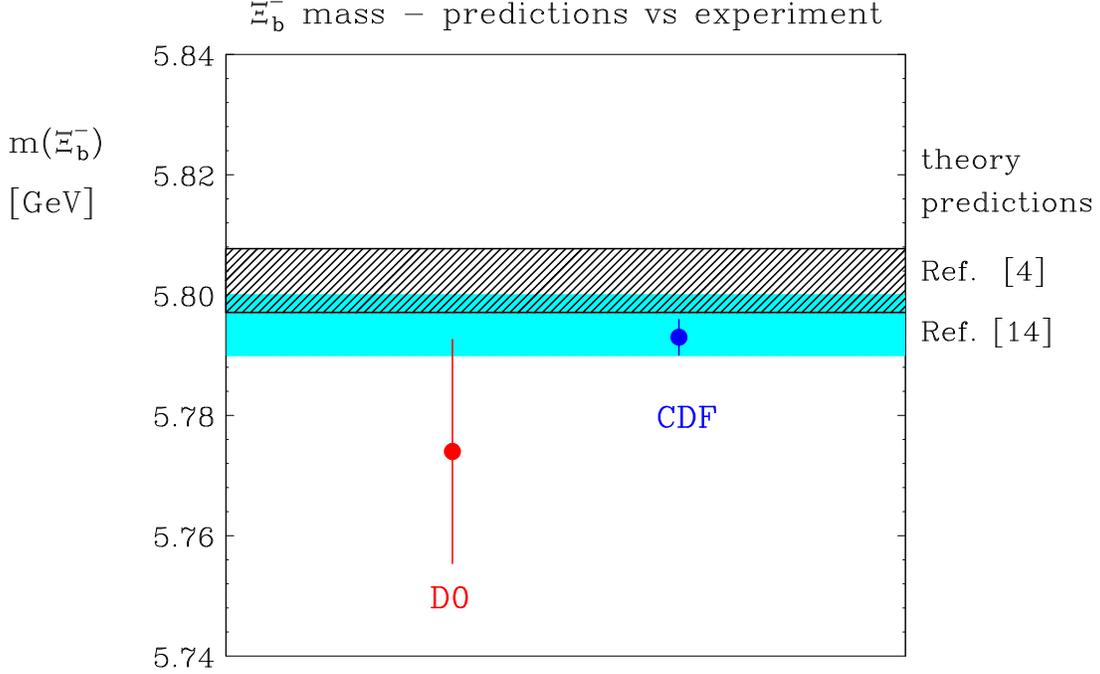}
\end{center}
\caption{Comparison of theoretical predictions and experimental results for the
$\Xi^-_b$ mass from D0 \cite{Abazov:2007ub} and CDF \cite{Aaltonen:2007un}
(adapted from \cite{Litvintsev:2007}).  The theoretical predictions are denoted
by the two horizontal bands, corresponding to Refs.~\cite{Jenkins:1996de} and
\cite{v1}, respectively.
\label{fig:comp}}
\end{figure}
The spin-averaged mass of these two states can be expressed as 
\begin{eqnarray}
\frac{2\Xi_q^*+\Xi_q'}{3} = m_q + m_s + m_u
+ \frac{v\langle \delta(r_{us}) \rangle}{m_um_s}~,
\end{eqnarray}
and as for the $\Xi_b$ case, the following prediction can be given:
\begin{eqnarray}
\frac{2\Xi_b^*+\Xi_b'}{3}=\frac{2\Xi_c^*+\Xi_c'}{3} + (m_b - m_c) +
\frac{2\Xi_c^*+\Xi_c'-3\Xi_c}{12} 
\Bigg(\frac{ \langle \delta(r_{us}) \rangle_{\Xi_b}}
{\langle \delta(r_{us}) \rangle_{\Xi_c}}-1\Bigg)~.
\end{eqnarray}
The predictions obtained using the same methods described above are given in
Table \ref{tab_Xib_star}.  Here the effect of the HF correction is negligible,
so the difference between the spin averaged mass $(2\Xi_b^*+\Xi_b')/3$ and
$\Xi_b$ is roughly $150-160$ MeV.

\begin{table}
\caption{Predictions for the spin averaged $\Xi_b'$ and $\Xi_b^*$ masses
with various confining potentials and methods of obtaining the quark mass
difference $m_b-m_c$.
\label{tab_Xib_star}}
\begin{center}
\begin{tabular}{cccc} \hline \hline
$m_b-m_c =$ & $\Lambda_b-\Lambda_c$ 
            & ${\Sigma_b}-{\Sigma_c}$ 
            & ${B_s}-{D_s}$          \\  
&Eq.~(\ref{eq_lambda_b_lambda_c})&Eq.~(\ref{eq_sigma_b_sigma_c})
&Eq.~(\ref{eq_B_s_D_s})\\ \hline 
No HF correction     & $5956\pm3$ & $5954\pm3$ & $5948\pm3$ \\
Linear               & $5957\pm4$ & $5954\pm4$ & $5948\pm4$ \\
Coulomb              & $5965\pm3$ & $5962\pm3$  & $5956\pm3$ \\
Cornell              & $5958\pm3$ & $5955\pm3$ & $5949\pm3$ \\ \hline \hline
\end{tabular}
\end{center}
\end{table}
\end{subsection}

\begin{subsection}{$\Xi_b^*-\Xi_b'$}
This mass difference will be small due to the large mass of the $b$ quark:
\begin{equation}
\Xi_q^*-\Xi_q'=3v\Bigg(\frac{\langle \delta(r_{qs}) \rangle}{m_qm_s}
+\frac{\langle \delta(r_{qu}) \rangle}{m_qm_u}\Bigg)
\end{equation}
We can once again use the $\Xi_c$ hadron masses: 
\begin{eqnarray}
\frac{\Xi_b^*-\Xi_b'}{\Xi_c^*-\Xi_c'}=
\frac
{\displaystyle{3v\Bigg(
\frac{\langle \delta(r_{bs}) \rangle}{m_bm_s}
+\frac{\langle \delta(r_{bu}) \rangle}{m_bm_u}\Bigg)}}
{\displaystyle{3v\Bigg(
\frac{\langle \delta(r_{cs}) \rangle}{m_cm_s}
+\frac{\langle \delta(r_{cu}) \rangle}{m_cm_u}\Bigg)}}=
\frac{m_c}{m_b}
\frac{\displaystyle{\Bigg ( \langle \delta(r_{bs}) \rangle_{\Xi_b}
 + \frac{m_s}{m_u} \langle \delta(r_{bu}) \rangle_{\Xi_b} \Bigg)}}
     {\displaystyle{\Bigg ( \langle \delta(r_{cs}) \rangle_{\Xi_c}
 + \frac{m_s}{m_u} \langle \delta(r_{cu}) \rangle_{\Xi_c} \Bigg)}}
\end{eqnarray}
This expression is strongly dependent on the confinement model. In the results
given in Table \ref{tab_Xibstar_Xibprime} we have used $m_s/m_u=1.5\pm0.1$,
$m_b/m_c=2.95\pm0.2$.

\begin{table}
\caption{Predictions for $M(\Xi_b^*)-M(\Xi_b')$ with various confining
potentials.
\label{tab_Xibstar_Xibprime}}
\begin{center}
\begin{tabular}{cc} \hline \hline
                     & $\Xi_b^*-\Xi_b'$      \\ \hline 
No HF correction      & $24\pm2$  \\
Linear                & $28\pm6$ \\
Coulomb               & $36\pm7$  \\
Cornell               & $29\pm6$  \\ \hline \hline
\end{tabular}
\end{center}
\end{table}

\end{subsection}
\end{section}

In the context of $\Xi'_b$ and $\Xi_b^*$ masses it is worth mentioning
two elegant relations among bottom baryons \cite{Savage:1995dw} which 
incorporate the effects of $SU(3)_f$ breaking:
\beq \label{eqn:sav1}
\Sigma_b + \Omega_b - 2 \Xi'_b   =  0~,
\eeq
\beq \label{eqn:sav2}
(\Sigma_b^* - \Sigma_b) + (\Omega_b^* - \Omega_b) - 2 (\Xi_b^* - \Xi'_b) =  0~,
\eeq
where isospin averaging is implicit.

\begin{section}{Effect of light-quark spin mixing on $\Xi_b$ and $\Xi'_b$}

In estimates up to this point we have assumed that the light-quark spins
in $\Xi_b$ and $\Xi'_b$ are purely $S=0$ and $S=1$, respectively.  The
differing hyperfine interactions between the $b$ quark and nonstrange or
strange quarks leads to a small admixture of the opposite-$S$ state in each
mass eigenstate \cite{Maltman:1980er,Lipkin:1981,Rosner:1981yh,Rosner:1992qa}.
The effective hyperfine Hamiltonian may be written \cite{Rosner:1981yh,%
Rosner:1992qa}
\begin{eqnarray}
H_{\rm eff} &=& M_0 + \lambda(\sigma_u \cdot \sigma_s + \alpha \sigma_u \cdot
\sigma_b + \beta \sigma_s \cdot \sigma_b)~,
\end{eqnarray}
where $M_0$ is the sum of spin independent terms, $\lambda \sim 1/(m_u m_s)$, $\alpha = m_s/m_b$, and $\beta = m_u/m_b$.
The calculation of $M_{3/2}$ is straightforward, as the expectation value of
each $\sigma_i \cdot \sigma_j$ in the $J=3/2$ state is 1.
For the $J=1/2$ states one has to diagonalize the $2 \times 2$ matrix
\beq
{\cal M}_{1/2} = \left[ \begin{array}{c c}
 M_0 - 3 \lambda & \lambda \sqrt{3} (\beta - \alpha) \cr
\lambda \sqrt{3} (\beta - \alpha) & M_0 + \lambda (1 - 2 \alpha - 2 \beta)
\end{array} \right]~.
\eeq

The eigenvalues of $H_{\rm eff}$ are thus
\begin{eqnarray}
M_{3/2} & = & M_0 + \lambda(1 + \alpha + \beta)~,\\
M_{1/2,\pm} & = & M_0 + \lambda[ -(1 + \alpha + \beta) \nonumber \\
  & \pm & 2\lambda(1 + \alpha^2 + \beta^2 - \alpha - \beta - \alpha \beta)^{1/2}~.
\end{eqnarray}

In the absence of mixing $(\alpha = \beta)$ one would have $M_{3/2} =
M_0 + \lambda(1 + 2 \alpha)$, $M_{1/2,+} = M_0 + \lambda (1 - 4 \alpha)$,
and $M_{1/2,-} = M_0 - 3 \lambda$.  

To see the effect of mixing, we rewrite the expression for 
$M_{1/2,\pm}$,
\beq
M_{1/2,\pm} = M_0 - \lambda(1 + \alpha + \beta) \pm
 2\lambda\left[\left(1 - {\alpha + \beta\over 2}\right)^2
 + \frac{3}{4}(\alpha - \beta)^2 \right]^{1/2}
\eeq
The effect of the mixing is seen in the term $\frac{3}{4}(\alpha -
\beta)^2$. Expanding $M_{1/2,\pm}$ to second order in small $\alpha-\beta$, 
we obtain
\beq
M_{1/2,\pm} \approx \left(\hbox{terms without mixing}\right) 
\pm \lambda\cdot \frac{3(\alpha-\beta)^2/4}{1 -  (\alpha + \beta)/2}~.
\eeq
For $m_u = 363$ MeV, $m_s=538$ MeV, 
and $m_b = 4900$ MeV \cite{Gasiorowicz:1981jz},
one has $\alpha \simeq 0.11$, $\beta \simeq 0.07$, while the discussion in the
previous section implies $\lambda \simeq 40$ MeV [Eq.\
(\ref{eq_HF_correction})]. 
Hence the effect of mixing on our predictions is negligible, amounting to 
$\pm 0.04$ MeV.

Since we use the $\Xi_c$ and $\Xi_c^\prime$ masses as input for $\Xi_b$,
it is also important to check the mixing effects on the former. Since
$m_b/m_c \sim 3$, this amounts to changing in the expressions above
$\alpha \rightarrow 3 \alpha$,
$\beta \rightarrow 3 \beta$. The corresponding effect of mixing
on $\Xi_c$ and $\Xi_c^\prime$ is $\sim 0.5$ MeV, still
negligible.

\end{section}

\begin{section}{Isospin splittings of $\Xi_b$ states}
\label{sec_xib_isospin}
The $\Xi_b^0$ mass is expected to be measured by the CDF collaboration through
the channel $\Xi_b^0 \to \Xi_c^+ \pi^-$, where $\Xi_c^+ \to \Xi^- \pi^+ \pi^+$,
$\Xi^- \to \Lambda \pi^-$, and $\Lambda \to p \pi^-$ \cite{Litvintsev:2007}.

The source for the isospin splitting ($\Delta I $) is the difference in the
mass and charge of the $u$ and $d$ quarks. These differences affect the hadron
mass in four ways \cite{Rosner:1998zc}: they change the constituent quark
masses ($\Delta M = m_d - m_u$), the Coulomb interaction ($V^{EM}$), and the
spin-dependent interactions -- both magnetic and chromo-magnetic ($V^{spin}$).
One can obtain a prediction for the $\Xi_b$ isospin splitting by extrapolation
from the $\Xi$ data, which has similar structure as far as EM interactions are
concerned (note that for $\Xi_b$ there are no spin-dependent interactions
between the heavy quark and the $su$ diquark which is coupled to spin zero):
\begin{eqnarray}
\Delta I(\Xi^*)&=&\Delta M + \Bigg[V^{EM}_{ssd} - V^{EM}_{ssu}\Bigg]
 +2 \Bigg[V^{spin}_{ds} - V^{spin}_{us}\Bigg]= 3.20\pm0.68~\textnormal{MeV} \\
\Delta I(\Xi)&=&\Delta M + \Bigg[V^{EM}_{ssd} - V^{EM}_{ssu}\Bigg]
 -4 \Bigg[V^{spin}_{ds} - V^{spin}_{us}\Bigg]= 6.85\pm0.21~\textnormal{MeV} \\
\strut\kern-5em
\Rightarrow\Delta I(\Xi_b)&=&\Delta M + \Bigg[V^{EM}_{ssd} - V^{EM}_{ssu}\Bigg]
 -3 \Bigg[V^{spin}_{ds} - V^{spin}_{us}\Bigg] \nonumber \\
&=&\frac{2\Delta I(\Xi^*)+\Delta I(\Xi)}{3}
 +\frac{\Delta I(\Xi)-\Delta I(\Xi^*)}{2}
=\frac{\Delta I(\Xi^*)+5\Delta I(\Xi)}{6} \nonumber \\
&=&6.24\pm0.21~\textnormal{MeV}
\end{eqnarray}
With the observed value \cite{Aaltonen:2007un} $M(\Xi_b^-) = (5792.9 \pm 2.5
\pm 1.7)$ MeV (the error from the D0 experiment is considerably
larger \cite{Abazov:2007ub}) and this estimate, we predict
$M(\Xi_b^0) = 5786.7 \pm 3.0$ MeV.

Another option is to use $\Xi_c$, which has the same spin-dependent
interactions, as a starting point:
\begin{eqnarray}
\Delta I(\Xi_c)&=&\Delta M + \Bigg[V^{EM}_{csd} - V^{EM}_{csu}\Bigg] -3 \Bigg[V^{spin}_{ds} - V^{spin}_{us}\Bigg]= 3.1\pm0.5~\textnormal{MeV} \\
\Rightarrow\Delta I(\Xi_b)&=&\Delta M + \Bigg[V^{EM}_{ssd} - V^{EM}_{ssu}\Bigg] -3 \Bigg[V^{spin}_{ds} - V^{spin}_{us}\Bigg]\\ \nonumber
&=&\Delta I(\Xi_c)+\Bigg[V^{EM}_{ssd} - V^{EM}_{ssu}\Bigg]-\Bigg[V^{EM}_{csd} - V^{EM}_{csu}\Bigg]\\ \nonumber
&=&\Delta I(\Xi_c)+\frac{2\Delta I(\Xi^*)+\Delta I(\Xi)}{3}-\frac{2\Delta I(\Xi_c^*)+\Delta I(\Xi_c')+\Delta I(\Xi_c)}{4}\\ \nonumber
&=&6.4\pm1.6~\textnormal{MeV}
\end{eqnarray}
We summarize the isospin splittings which have been used in these calculations
in Table \ref{tab:split}.  All masses have been taken from the 2007 updated
tables of the Particle Data Group \cite{PDG07}, and all values of $\Delta I$
are defined as $M$(baryon with $d$ quark) - $M$(baryon with $u$ quark).

\begin{table}
\caption{Isospin splittings $\Delta I$ used in calculating
$\Delta I(\Xi_b) \equiv M(\Xi_b^-) - M(\Xi_b^0)$.
\label{tab:split}}
\begin{center}
\begin{tabular}{l l} \hline \hline
Splitting & Value (MeV) \\ \hline
$\Delta I(\Xi)$     & $\phantom{-}6.85 \pm 0.21$ \\
$\Delta I(\Xi^*)$   & $\phantom{-}3.20 \pm 0.68$ \\
$\Delta I(\Xi_c)$   &  $\phantom{-}3.1\phantom{0} \pm 0.5$  \\
$\Delta I(\Xi'_c)$  &  $\phantom{-}2.3\phantom{0} \pm 4.24$ \\
$\Delta I(\Xi^*_c)$ & ${-}0.5\phantom{0} \pm 1.84$ \\ \hline \hline
\end{tabular}
\end{center}
\end{table}

\end{section}

\begin{section}{$\Lambda_b$ and $\Xi_b$ orbital excitations}
\strut\vskip-2em 
\begin{table}[h]
\caption{Masses of $\Lambda$ and $\Xi$ baryon ground states and orbital
excitations~\cite{PDG07}.
\label{tab:orbital}}
\begin{center}
\begin{tabular}{l c c c c} \hline \hline
         & $\Lambda$              &$\Lambda_c$         &$\Xi_c^+$           &
$\Xi_c^0$           \\ \hline
$M(1/2^+)$ & $1115.683 \pm 0.006$  & $2286.46 \pm 0.14$ &$2467.9  \pm 0.4$   &
$2471.0   \pm 0.4$  \\
$M(1/2^-)$ & $1406.5   \pm 4.0$    & $2595.4  \pm 0.6$  &$2789.2  \pm 3.2 $  &
$2791.9   \pm 3.3$  \\
$M(3/2^-)$ & $1519.5   \pm 1.0$    & $2628.1  \pm 0.6$  &$2816.5  \pm 1.2 $  &
$2818.2   \pm 2.1$  \\ \hline
\hline
\end{tabular}
\end{center}
\end{table}

In the heavy quark limit, the ($1/2^-$) and ($3/2^-$) $\Lambda^*$ and $\Xi^*$
excitations listed in Table \ref{tab:orbital}
can be interpreted as a P-wave isospin-0 spinless diquark coupled to the heavy
quark.  Under this assumption, the difference between the spin averaged mass
of the $\Lambda^*$ baryons and the ground state $\Lambda$ is only the orbital
excitation energy of the diquark.
\vbox{
\begin{eqnarray}
\label{eqn:lex}
\Delta E_L(\Lambda) \equiv
\frac{2\Lambda^*_{[3/2]} +\Lambda^*_{[1/2]}}{3}  -\Lambda    &=&366.15\pm1.49~\textnormal{MeV} \nonumber \\
\Delta E_L(\Lambda_c) \equiv
\frac{2\Lambda^*_{c[3/2]}+\Lambda^*_{c[1/2]}}{3} -\Lambda_c  &=&330.74\pm0.47~\textnormal{MeV}  \\
\Delta E_L(\Xi_c) \equiv
\frac{2\Xi^{*}_{c[3/2]}  +\Xi^{*}_{c[1/2]}}{3}  -\Xi_c &=&339.11\pm1.11~\textnormal{MeV} \nonumber
\end{eqnarray}
}

The spin-orbit splitting seems to behave like $1/m_Q$:
\begin{eqnarray}
\label{eqn:fs}
\Lambda^*_{[3/2]}  -\Lambda^*_{[1/2]} &=&113.0\pm4.1 ~\textnormal{MeV} \nonumber \\
\Lambda^*_{c[3/2]} -\Lambda^*_{c[1/2]}&=&\phantom{1}32.7 \pm0.8~\textnormal{MeV}  \\
\Xi^*    _{c[3/2]} -\Xi^*    _{c[1/2]}&=&\phantom{1}26.9\pm2.6~\textnormal{MeV} \nonumber
\end{eqnarray}
where the $\Xi_c$ entries are isospin averages.

The orbital excitation energies in Eq.\ (\ref{eqn:lex}) may be extrapolated to
the case of excited $\Lambda_b$ baryons in the following manner.  Energy
spacings in a power-law potential $V(r) \sim r^\nu$ behave with reduced mass
$\mu$ as $\Delta E \sim \mu^p$, where $p = - \nu/(2+\nu)$ \cite{Quigg:1977dd}.
For light quarks in the confinement regime, one expects $\nu = 1$ and $p =
-1/3$, while for the $c \bar c$ and $b \bar b$ quarkonium states, with nearly
equal level spacings, an effective power is $\nu \simeq 0$ and $p \simeq 0$.
One should thus expect orbital excitations to scale with some power $-1/3
\le p \le 0$.  One can narrow this range by comparing the $\Lambda$
and $\Lambda_c$ excitation energies and estimating $p$ with the help of
reduced masses $\mu$ for the $\Lambda$ and $\Lambda_c$.

\bea
\frac{\mu(\Lambda_c)}{\mu(\Lambda)} =
\frac{M[ud]~m_c}{M[ud] + m_c}\frac{M[ud] + m_s}{M[ud]~m_s}=
\frac{M(\Lambda)}{M(\Lambda_c)}\frac{m_c}{m_s}=1.55
\eea
Now we use the ratio $\Delta E_L(\Lambda_c)/\Delta E_L(\Lambda) = 0.903 \pm
0.004$ to extract an effective power $p = -0.23 \pm 0.01$ which will be used
to extrapolate to the $\Lambda_b$ system:
\bea
\Delta E_L(\Lambda_b) &=&
\Delta E_L(\Lambda_c) \left[ \frac{\mu(\Lambda_b)}{\mu(\Lambda_c)} \right]^p
=\Delta E_L(\Lambda_c)
\left[ \frac{M(\Lambda_c)}{M(\Lambda_b)}\frac{m_b}{m_c} \right]^p
\strut\nonumber\\
\strut\nonumber\\
&=&\Delta E_L(\Lambda_c)
\left[ \frac{M(\Lambda_c)[M(\Lambda_b)-M(\Lambda)+m_s]}
{M(\Lambda_b)[M(\Lambda_c)-M(\Lambda)+m_s]} \right]^p
\\
\strut\nonumber\\
&=&
\Delta E_L(\Lambda_c)
\left[ \frac{\displaystyle 1-{M(\Lambda)- m_s\over M(\Lambda_b)}}
{\displaystyle 1-{M(\Lambda)- m_s\over M(\Lambda_c)}} \right]^p
 = 317 \pm 1~{\rm MeV}
\nonumber
\label{delta-E-L-Lambda-c}
\eea
where the last form of the expression shows the explicit dependence of the
result on $m_s$.
Using the value $M(\Lambda_b) = (5619.7 \pm 1.2 \pm 1.2)$ MeV observed by
the CDF Collaboration \cite{Acosta:2005mq}, and rescaling the fine-structure
splittings of Eq.\ (\ref{eqn:fs}) by $1/m_Q$ with $m_b/m_c = 2.95 \pm 0.06$,
we find

\begin{equation}
M(\Lambda^*_{b[3/2]}) - M(\Lambda^*_{b[1/2]}) = \frac{m_c}{m_b} (M(\Lambda^*_{c[3/2]})
 - M(\Lambda^*_{c[1/2]})) = (11.1 \pm 0.4)~{\rm MeV}~,
\end{equation}
\begin{equation}
M(\Lambda^*_{b[1/2]}) = (5929 \pm 2)~{\rm MeV}~,~~
M(\Lambda^*_{b[3/2]}) = (5940 \pm 2)~{\rm MeV}~.
\end{equation}

The observed values of the $\Sigma_b$ masses \cite{Aaltonen:2007rw},
\begin{eqnarray}
M(\Sigma_b^-) &=& 5815.2{\pm}1.0 ({\rm stat.}) \pm 1.7 ({\rm syst.})
~{\rm MeV} \nonumber \\
M(\Sigma_b^+) &=& {5807.8\,}^{+2.0}_{{-}2.2}\,\,({\rm stat.})
\strut\pm\strut 1.7 ({\rm syst.})
~{\rm MeV}
\end{eqnarray}
are sufficiently close to the predicted values of $M(\Lambda^*_{b[1/2,3/2]})$
that the decays
\break
 $\Lambda^*_{b[1/2,3/2]} \to \Sigma_b^\pm \pi^\mp$ \ are
forbidden.  The $\Lambda^*_{b[1/2,3/2]}$ should decay directly to $\Lambda_b
\pi^+ \pi^-$.
\downstrut 

A similar calculation may be performed for the orbitally-excited $\Xi_b$
states.  Here, to a good approximation, one may
regard the $[sd]$ diquark in $\Xi_b^-$ or the $[su]$ diquark in $\Xi_b^0$ as
having spin zero, so that methods similar to those applied for excited
$\Lambda_b$ states should be satisfactory.  We find
\beq
\Delta E_L(\Xi_b)
= \Delta E_L(\Xi_c) \left[ \frac{\mu(\Xi_b)}{\mu(\Xi_c)} \right]^p
=\Delta E_L(\Xi_c) \left[ \frac{M(\Xi_c)}{M(\Xi_b)}\frac{m_b}{m_c} \right]^p
= (322 \pm 2)~{\rm MeV}~.
\label{Delta-E-L-Xi-b}
\eeq
Now we use the observed $\Xi_b^-$ mass \cite{Aaltonen:2007un} $M(\Xi_b^-) =
(5792.9 \pm 2.5 \pm 1.7)$ MeV
and our estimate of isospin splitting
$M(\Xi_b^-) - M(\Xi_b^0) = 6.4 \pm 1.6$ MeV to predict the isospin-averaged
value $M(\Xi_b) = 5790 \pm 3$ MeV. We then rescale the fine-structure splitting
(\ref{eqn:fs}) and find
\begin{equation}
\Xi^*_{b[3/2]} - \Xi^*_{b[1/2]} = \frac{m_c}{m_b} (\Xi^*_{c[3/2]}
 - \Xi^*_{c[1/2]}) = (9.1 \pm 0.9)~{\rm MeV}~,
\end{equation}
\begin{equation}
M(\Xi^*_{b[1/2]}) = (6106 \pm 4)~{\rm MeV}~,~~
M(\Xi^*_{b[3/2]}) = (6115 \pm 4)~{\rm MeV}~.
\end{equation}
The lower state decays to $\Xi_b \pi$ via an S-wave, while the higher
state decays to $\Xi_b \pi$ via a D-wave, and hence should be narrower.
Decays to $\Xi'_b \pi$ and $\Xi^*_b \pi$ also appear to be just barely
allowed, given the values of $M(\Xi'_b,\Xi^*_b)$ predicted here.
\end{section}

\begin{section}{$\Omega_b$ mass prediction}

Taking the approach implemented in Sec.\ 3 for the prediction
of the $\Xi_b$ mass, the spin averaged mass of $\Omega_b$ can be obtained by
extrapolation from available data for $\Omega_c$ and a correction based on
strange meson masses, as listed in Table \ref{tab:masses}:
\begin{eqnarray}
\label{Omegab-spin-ave}
M(\widetilde{\Omega_b})&\equiv&
\frac{2M(\Omega_b^*)+M(\Omega_b)}{3}=\frac{2M(\Omega_c^*)+M(\Omega_c)}{3}
+{(m_b-m_c)\medstrut}_{B_s-D_s}
\downstrut 
\\
 \nonumber
&=&\frac{2M(\Omega_c^*)+M(\Omega_c)}{3}
+\frac{3M(B_s^*)+M(B_s)}{4}-\frac{3M(D_s^*)+M(D_s)}{4}
\downstrut 
\\
\nonumber
&=&6068.9\pm 2.4~\textnormal{MeV}
\end{eqnarray}
where $M(\widetilde{X})$ denotes the spin-averaged mass
 that cancels out the hyperfine interaction between the
heavy quark and the diquark containing lighter quarks.
\downstrut

The HF splitting can be estimated as follows:
\begin{eqnarray}
M(\Omega_b^*)-M(\Omega_b) & = &(M(\Omega_c^*)-M(\Omega_c))\frac{m_c}{m_b} =
(24.0 \pm 0.7)~\textnormal{MeV}~,
\end{eqnarray}
where we have used the experimental mass difference
\cite{Aubert:2006je} $M(\Omega_c^*) - M(\Omega_c) =\break
(70.8 \pm 1.0 \pm 1.1)~{\rm MeV} = (70.8 \pm 1.5)~{\rm MeV}$
with $m_b/m_c$ taken to be $2.95 \pm 0.06 $, as discussed in the Appendix.
This gives the following mass predictions:
\begin{equation}
\Omega_b^*=(6076.9\pm 2.4)~\rm{MeV}; ~ ~ ~ \Omega_b=(6052.9\pm2.4)~\rm{MeV}~.
\end{equation}
Taking into account the wavefunction correction as described in
\cite{Keren-Zur:2007vp}, one must add the following correction to
the spin averaged mass:
\begin{eqnarray}
v\Bigg[\frac{\langle \delta(r_{ss}) \rangle_{\Omega_b}}{m_s^2}
-\frac{\langle \delta(r_{ss}) \rangle_{\Omega_c}}{m_s^2}\Bigg]
&=& v\frac{\langle \delta(r_{ss}) \rangle_{\Omega_c}}{m_s^2}
\Bigg[\frac{\langle \delta(r_{ss}) \rangle_{\Omega_b}}{\langle \delta(r_{ss})
\rangle_{\Omega_c}}-1\Bigg] \nonumber \\
&\approx&(50\pm10)\Bigg[\frac{\langle \delta(r_{ss}) \rangle_{\Omega_b}}
{\langle \delta(r_{ss}) \rangle_{\Omega_c}}-1\Bigg]=(2.0\pm1.1)~\rm{MeV}
\label{WF-corr}
\end{eqnarray}
where the contact probability ratio was computed using variational methods
\beq
\frac{\langle \delta(r_{ss}) \rangle_{\Omega_b}}{\langle \delta(r_{ss}) \rangle_{\Omega_c}}=1.04\pm0.02~,
\eeq
and we used the following calculation to evaluate the strength of the $ss$ HF
interaction:
\begin{eqnarray}
50~\rm{MeV}&\approx&M(\Omega)+\frac{1}{4}(2M(\Xi_c^*)+M(\Xi_c')+M(\Xi_c))
\nonumber
\\
&&-\frac{1}{3}(2M(\Xi^*)+M(\Xi))-\frac{1}{3}(2M(\Omega_c^*)+M(\Omega_c))
=\nonumber \\
&=&\Bigg(3m_s+3v\frac{\langle \delta(r_{ss})
\rangle_{\Omega}}{m_s^2}\Bigg)+\Bigg(m_u+m_s+m_c\Bigg) \nonumber \\
&&-\Bigg(2m_s+m_u+v\frac{\langle \delta(r_{ss}) \rangle_{\Xi}}{m_s^2}\Bigg)
-\Bigg(2m_s+m_c+v\frac{\langle \delta(r_{ss}) \rangle_{\Omega_c}}{m_s^2}\Bigg)
\nonumber\\
&\approx& v\frac{\langle \delta(r_{ss}) \rangle}{m_s^2}
\end{eqnarray}

\begin{table}
\caption{Hadron masses used in the calculation of the $\Omega_b$ mass
prediction
\label{tab:masses}}
\begin{center}
\begin{tabular}{l c} \hline \hline
Splitting                             & Value (MeV) \\ \hline
$M(\Omega_c)$                         & $2697.5\phantom{0}\pm 2.6$ \\
$M(\Omega_c^*)$                       & $2768.3\phantom{0}\pm 3.0$ \\
$M(\Omega_c^*)-M(\Omega_c)$           & $\phantom{00}70.8\phantom{0}\pm 1.5$ \\
$M(D_s)$                              & $1968.49 \pm 0.34$ \\
$M(D_s^*)$                            & $2112.3\phantom{0}\pm 0.5$ \\
$M(B_s)$                              & $5366.1\phantom{0}\pm 0.6$ \\
$M(B_s^*)$                            & $5412.0\phantom{0}\pm 1.2$ \\
$M(B_s^*)-M(B_s)$                     & $\phantom{00}45.9\phantom{0} \pm 1.2$ \\$M(\Xi_c^0)$                          & $2471.0\phantom{0} \pm 0.4$ \\
$M(\Xi_b^-)$                          & $5792.9\phantom{0} \pm 3.0$ \\
\hline
\hline
\end{tabular}
\end{center}
\end{table}

\subsubsection*{An alternate derivation of the $\Omega_b$ mass from the
$\Xi_b - \Xi_c$ mass difference}

Thanks to new measurements of the $\Xi_b^-$ mass \cite{Abazov:2007ub,%
Aaltonen:2007un}, we now have another way to estimate the spin-averaged
$\Omega_b$ mass.  Following the approach in previous sections,
the $\Xi_b^- - \Xi_c^0$ mass difference can be schematically written as
\beq
\begin{array}{ccccccc}
M(\Xi_b^-)-M(\Xi_c^0)&=& (m_b - m_c) &+& \hbox{(wavefunction correction)}
&+& \hbox{(EM correction)}
\\
\\
&=& (m_b - m_c) &+& ({-}4\pm4)~{\rm MeV}
&+&(V^{EM}_{bsd}-V^{EM}_{csd})
\\
\\
\end{array}
\eeq
where the value of the wave function correction is calculated as described in
Sec.\ 3, and the last term denotes the EM interactions of
the relevant quarks.

Similarly, the spin-averaged $\Omega_b-\Omega_c$ mass difference can be
written as
\beq
\begin{array}{ccccccc}
M(\widetilde{\Omega_b})-M(\widetilde{\Omega_c})
&=& (m_b - m_c) &+& \hbox{(wavefunction correction)}
&+& \hbox{(EM correction)}
\\
\\
&=& (m_b - m_c) &+& (2.0\pm1.1)~{\rm MeV}
&+&(V^{EM}_{bss}-V^{EM}_{css})
\\
\\
\end{array}
\eeq
where the wave-function correction is given in Eq.~\eqref{WF-corr}.
\downstrut

Since the $b$ and $s$ quarks have the same charge, the EM contribution
\hbox{$V^{EM}_{bss}-V^{EM}_{css}$} to the $\Omega_b - \Omega_c$ mass difference
is almost the same as the EM contribution \hbox{$V^{EM}_{bsd}-V^{EM}_{csd}$}
to the $\Xi_b^- - \Xi_c^0$ mass difference, modulo a negligible correction from
the change in the mean radius of the relevant baryons. We then immediately
obtain
\beq
M(\widetilde{\Omega_b})-M(\widetilde{\Omega_c}) =
M(\Xi_b^-)-M(\Xi_c^0)+(6.0\pm4.1)~{\rm MeV}
\eeq
which leads to
\beq
M(\widetilde{\Omega_b})=6072.6\pm 5.6~\textnormal{MeV}
\label{Omegab-from-Xi}
\eeq
to be compared with
$M(\widetilde{\Omega_b})=6070.9\pm 2.7~\textnormal{MeV}$
from Eqs.~\eqref{Omegab-spin-ave} and \eqref{WF-corr}.
\downstrut

The consistency of these two estimates, based on different
experimental inputs, is a strong indication that
both the central values and the error estimates are reliable.
Moreover, the estimate in Eq.~\eqref{Omegab-from-Xi} includes EM corrections,
while the estimate Eqs.~\eqref{Omegab-spin-ave} does not, thus indicating
that the EM corrections are likely to be smaller than our error estimate.
Consequently, in the following we use the estimate
\eqref{Omegab-from-Xi}.

\subsubsection*{Wave function correction to the hyperfine splitting}

We must also compute the correction to the HF splitting
\begin{eqnarray}
M(\Omega_b^*)-M(\Omega_b)&=&
(M(\Omega_c^*)-M(\Omega_c))\frac{m_c}{m_b}\frac{\langle \delta(r_{bs})
\rangle_{\Omega_b}}{\langle \delta(r_{cs}) \rangle_{\Omega_c}}
=30.7 \pm1.3~\textnormal{MeV}
\end{eqnarray}
where we used
\beq
\frac{\langle \delta(r_{bs}) \rangle_{\Omega_b}}{\langle \delta(r_{cs})
 \rangle_{\Omega_c}}=1.28\pm0.04~,
\eeq
leading to the following predictions:
\begin{equation}
\Omega_b^* = 6082.8\pm5.6~\rm{MeV}; ~ ~ ~ \Omega_b = 6052.1\pm5.6~\rm{MeV}
\label{omegab_pred}
\end{equation}

\subsubsection*{An alternative derivation of HF splitting from effective supersymmetry}
An alternative approach to estimate the HF splitting is to use the effective
meson-baryon supersymmetry discussed in \cite{Karliner:2006ny} and
apply it to the case of hadrons related by changing a strange antiquark
$\bar s$ to a doubly strange $ss$ diquark coupled to spin $S = 1$:

\begin{equation}
\begin{array}{ccccccc}
\displaystyle
{{M(\Omega_b^*) - M(\Omega_b)}\over{M(B_s^*)-M(B_s)}} &=&
\displaystyle
{{M(\Omega_c^*){-}M(\Omega_c)}\over{M(D_s^*){-}M(D_s)}}
&=&
\displaystyle
{{M(\Xi^*){-}M(\Xi)}\over{M(K^*){-}M(K)}}
\strut\kern-5em\strut
\label{eq:newpredo}
\\
\\
[4pt]
&\approx &
0.49 \pm 0.01
&\approx&
0.54
\end{array}
\end{equation}

\begin{eqnarray}
\Omega_b^*-\Omega_b=(B_s^*-B_s)(0.52\pm0.02)=23.9\pm1.1~\textnormal{MeV}
\end{eqnarray}
This gives \begin{equation}
\Omega_b^* = 6080.6\pm5.6~\rm{MeV};
~ ~ ~
\Omega_b = 6056.7\pm5.6~\rm{MeV}~.
\end{equation}

\end{section}

\begin{section}{Comparisons with other approaches}

We begin by comparing $M(\Sigma^*_b) - M(\Sigma_b) = 20.0 \pm 0.3$ MeV as
predicted in Sec.\ 2 with other predictions and data.  The first of Refs.\
\cite{Jenkins:1996de} finds $M(\Sigma_b^*) - M(\Sigma_b) = 23.8$ MeV, the
second finds 15.8 MeV, Ref.\ \cite{Ebert:2005xj} finds 29 MeV, and Ref.\
\cite{Liu:2007fg} finds $26 \pm 1$ MeV.  The experimental value is $21.2^{+2.0}
_{-1.9}$ MeV \cite{Aaltonen:2007rw}.  A recent analysis \cite{Jenkins:2007dm}
uses $M(\Sigma^*_b) - M(\Sigma_b)$ as an input to a sum rule
\beq
M(\Sigma_b^*) - M(\Sigma_b) - 2[M(\Xi^*_b) - M(\Xi_b)]
+ M(\Omega^*_b) - M(\Omega_b) = \pm 0.28~{\rm MeV}~.
\eeq 
Our predictions entail a value of ${-}7 \pm 12$ MeV for the right hand side.
The deviation between these two predictions is significant because they arise
from a difference in the sign between the SU(3) breaking contributions.

The sign in our prediction
\begin{equation}
M(\Sigma_b^*) - M(\Sigma_b) < M(\Omega_b^*) - M(\Omega_b)
\end {equation}
appears to be counterintuitive, since the color hyperfine
interaction  is inversely proportional to the quark mass. The expectation
value of the interaction with the same wave function for $\Sigma_b$ and
$\Omega_b$ violates our
inequality.  When wave function effects are included, the inequality
is still violated if the potential is linear, but is satisfied in
predictions which use the Cornell potential \cite{Keren-Zur:2007vp}.

This reversed inequality is  not predicted by other recent approaches
\cite{Ebert:2005xj,Roberts:2007ni,Jenkins:2007dm}  which all predict an 
$\Omega_b$ splitting smaller than a $\Sigma_b$  splitting. 

However the reversed inequality is also seen in the 
corresponding charm experimental data,
\bea
M(\Sigma_c^*) - M(\Sigma_c) & < & M(\Omega_c^*) - M(\Omega_c) 
\nonumber
\\
64.3 \pm 0.5 \hbox{MeV}\strut\kern0.5em\strut  &&
\strut\kern0.2em\strut  
 70.8\pm 1.5 \hbox{MeV}
\nonumber\\
\eea

This suggests that the sign of the $SU(3)$ symmetry breaking gives information 
about the form of the potential. It is of interest to follow this clue
theoretically and experimentally. 

We compare our results with some other recent approaches \cite{Ebert:2005xj,%
Roberts:2007ni,Jenkins:2007dm} and with data in Table \ref{tab:comp}.  The
results of Ref.\ \cite{Liu:2007fg}, based on Heavy Quark Effective Theory
and QCD sum rules, typically carry $\pm 80$ MeV errors so we omit them here.
We also take note of a very recent set of predictions which differ
substantially from those in Table \ref{tab:comp} \cite{Gerasyuta:2008zy}.
The main difference between our predictions for $\Xi_b$ and $\Omega_b$ states
and other recent ones \cite{Ebert:2005xj,Liu:2007fg,Roberts:2007ni,Jenkins:%
2007dm} is the use of masses of hadrons containing strange quarks
to obtain the quark mass difference $m_b-m_c$.  We also take into
account wave function corrections, particularly important for
the hyperfine splitting between $\Omega_b^*$ and $\Omega_b$.

\end{section}

\begin{section}{Summary}

We have predicted the masses of several baryons containing $b$ quarks, using
descriptions of the color hyperfine interaction which have proved successful
for earlier predictions.  Correcting for wave function effects, we have shown
that predictions for $M(\Xi_b)$ based on the masses of $\Xi_c$, $\Xi'_c$, and
$\Xi^*_c$ lie in the range of 5790 to 5800 MeV, depending on how $m_b - m_c$ is
estimated.  Wave function differences tend to affect these predictions by only
a few MeV.  The spin-averaged mass of the states $\Xi'_b$ and $\Xi^*_b$ is
predicted to lie  around 150 to 160 MeV above $M(\Xi_b)$, while the hyperfine
splitting between $\Xi'_b$ and $\Xi^*_b$ is predicted to lie in the rough range
of 20 to 30 MeV.

\begin{table}
\caption{Comparison of predictions for $b$ baryons with those of some other
recent approaches \cite{Ebert:2005xj,Roberts:2007ni,Jenkins:2007dm} and with
experiment.  Masses quoted are isospin averages unless otherwise noted.
Our predictions are those based on the Cornell potential.
\label{tab:comp}}
\begin{center}
\begin{tabular}{cccccc} \hline \hline
         & \multicolumn{5}{c}{Value in MeV} \\
Quantity & Refs.\ \cite{Ebert:2005xj} & Ref.\ \cite{Roberts:2007ni} &
Ref.\ \cite{Jenkins:2007dm} & This work & Experiment \\ \hline
$M(\Lambda_b)$ & 5622 & 5612 & Input & Input & 5619.7$\pm$1.7 \\
$M(\Sigma_b)$ & 5805 & 5833 & Input & -- & 5811.5$\pm$2 \\
$M(\Sigma^*_b)$ & 5834 & 5858 & Input & -- & 5832.7$\pm$2 \\
$M(\Sigma^*_b) - M(\Sigma_b)$ & 29 & 25 & Input & 20.0$\pm$0.3
 & 21.2$^{+2.2}_{-2.1}$ \\
$M(\Xi_b)$ & 5812 & 5806$^a$ & Input & 5790--5800 & 5792.9$\pm$3.0$^b$ \\
$M(\Xi'_b)$  & 5937 & 5970$^a$ & 5929.7$\pm$4.4 & 5930$\pm$5 & -- \\
$\Delta M(\Xi^b)^c$ & -- & -- & -- & 6.4$\pm$1.6 & -- \\
$M(\Xi^*_b)$ & 5963 & 5980$^a$ & 5950.3$\pm$4.2 & 5959$\pm$4 & -- \\
$M(\Xi^*_b) - M(\Xi'_b)$ & 26 & 10$^a$ & 20.6$\pm$1.9 & 29$\pm$6 & -- \\
$M(\Omega_b)$ & 6065 & 6081 & 6039.1$\pm$8.3 & 6052.1$\pm$5.6 & --\\
$M(\Omega_b^*)$ & 6088 & 6102 & 6058.9$\pm$8.1 & 6082.8$\pm$5.6 & --\\
$M(\Omega_b^*) - M(\Omega_b)$ & 23 & 21 & 19.8$\pm$3.1 & 30.7$\pm$1.3 & -- \\
$M(\Lambda^*_{b[1/2]})$ & 5930 & 5939 & -- & $5929\phantom{.0} \pm 2$ & --\\
$M(\Lambda^*_{b[3/2]})$ & 5947 & 5941 & -- & $5940\phantom{.0} \pm 2$ & --\\
$M(\Xi^*_{b[1/2]})$ & 6119 & 6090 & -- & $6106\phantom{.0} \pm 4$ & --\\
$M(\Xi^*_{b[3/2]})$ & 6130 & 6093 & -- & $6115\phantom{.0} \pm 4$ & --\\
\hline \hline
\end{tabular}
\end{center}
\leftline{$^a$Value with configuration mixing taken into account; slightly
higher without mixing.}
\leftline{$^b$CDF \cite{Aaltonen:2007un} value of $M(\Xi_b^-)$.}
\leftline{$^c$$M$(state with $d$ quark) -- $M$(state with $u$ quark).}
\end{table}

We have evaluated the isospin splitting of the $\Xi_b$ states and find
$\Delta I (\Xi_b) \equiv M(\Xi_b^-) - M(\Xi_b^0) = 6.24 \pm 0.21$ MeV on the
basis of an extrapolation from the $\Xi$ and $\Xi^*$ states.  This value is
consistent with one which includes information from the $\Xi_c$ states,
$\Delta I (\Xi_b) = 6.4 \pm 1.6$ MeV.

We predict $M(\Omega_b) = 6052.1 \pm 5.6$ MeV and $M(\Omega^*_b) = 6082.8 \pm
5.6$ MeV.  These values differ from some others which have appeared in recent
literature because we use hadrons containing strange quarks to evaluate the
effective $b-c$ mass difference, include electromagnetic
contributions, and employ different hyperfine splittings.

We have also evaluated the orbital excitation energy for $\Lambda_b$ and
$\Xi_b$ states in which the light diquark ($ud$ or $us$) remains in a
state of $L=S=0$.  Precise predictions have been given for the masses
of the states $\Lambda^*_{b[1/2,3/2]}$ and $\Xi^*_{b[1/2,3/2]}$.
\medskip

We look forward to tests of some of the predictions summarized in Table
\ref{tab:comp} in experiments at the Fermilab Tevatron
and the CERN Large Hadron Collider.  

\end{section}
\section*{Acknowledgements}
J.L.R. wishes to acknowledge the hospitality of Tel Aviv University during
the early stages of this investigation.  Part of this work was performed while
J.L.R. was at the Aspen Center for Physics.  We thank Dmitry Litvintsev for
providing his figure comparing theoretical predictions with measurements of the
$\Xi^-_b$ mass.  This research was supported in part by a grant from Israel
Science Foundation administered by Israel Academy of Science and Humanities.
The research of H.J.L. was supported in part by the U.S. Department of Energy,
Division of High Energy Physics, Contract DE-AC02-06CH11357.
The work of J.L.R. was supported by the U.S. Department of Energy, Division of
High Energy Physics, Grant No.\ DE-FG02-90ER40560.

\end{document}